\documentclass[prl,twocolumn,superscriptaddress]{revtex4-2}
\usepackage{bm}

\usepackage{qcircuit}
\usepackage{graphicx}
\usepackage{dcolumn}
\usepackage{braket}
\usepackage{amsmath}
\usepackage{color}
\usepackage{amssymb}
\usepackage{amsthm}
\usepackage{amsmath}
\usepackage{dsfont}
\usepackage{adjustbox}
\usepackage{tikz}
\usetikzlibrary{decorations.markings}
\usepackage{graphicx}
\usepackage{subcaption}
\usepackage{pgfplots}
\pgfplotsset{compat=1.3}
\usepgfplotslibrary{fillbetween}
\usetikzlibrary{patterns}
\usepackage{mathtools}
\usepackage{xcolor}
\usepackage{enumerate}
\usepackage{soul}
\usepackage[many]{tcolorbox}    	
\usepackage{physics}

\usepackage[labelfont=bf,
   justification=Justified,
   format=plain]{caption}
\usepackage[hidelinks]{hyperref}
\hypersetup{
    colorlinks,
    linkcolor={red!50!black},
    citecolor={blue!50!black},
    urlcolor={blue!80!black}
}

\definecolor{color3}{RGB}{94, 201, 98}
\definecolor{color2}{RGB}{33, 145, 140}
\definecolor{color1}{RGB}{59, 82, 139}

\definecolor{plasma_color3}{RGB}{248, 149, 64}
\definecolor{plasma_color2}{RGB}{204, 71, 120}
\definecolor{plasma_color1}{RGB}{126, 3, 168}

\definecolor{colore}{rgb}{0.9, 0.9, 0.9}
\newtcolorbox{boxA}{
    colback = colore, 
    boxrule = 0pt  
}

\begin{document}

\title{Exponential optimization of adiabatic quantum-state preparation}

\author{Davide Cugini}
\affiliation{Dipartimento di Fisica, Universit\`a di Pavia, via Bassi 6, 27100,  Pavia, Italy}

\author{Davide Nigro}
\affiliation{Dipartimento di Fisica, Universit\`a di Pavia, via Bassi 6, 27100,  Pavia, Italy}

\author{Mattia Bruno}
\affiliation{Dipartimento di Fisica, Universit\`a degli Studi di Milano--Bicocca, and INFN, sezione di Milano--Bicocca, Piazza della Scienza 3, I-20126 Milano, Italy}

\author{Dario Gerace}
\affiliation{Dipartimento di Fisica, Universit\`a di Pavia, via Bassi 6, 27100,  Pavia, Italy}

\begin{abstract}
The preparation of a given quantum state 
on a quantum computing register is a typically demanding operation, requiring a number of elementary gates that scales exponentially with the size of the problem. 
Using the adiabatic theorem for state preparation, whose error decreases exponentially as a function of the preparation time, we derive an explicit analytic expression for the dependence of the characteristic time on the Hamiltonian used in the adiabatic evolution.
Exploiting this knowledge, we then design a preconditioning term that modifies the adiabatic preparation, thus reducing its characteristic time
and hence giving an exponential advantage in state preparation.
We prove the efficiency of our method with extensive numerical experiments on prototypical spin-models, which gives a promising strategy to perform quantum simulations of manybody models via Trotter evolution on near-term quantum processors. 
\end{abstract}

\maketitle

\textit{Introduction}.--
We are currently living the so-called Noisy Intermediate-Scale Quantum (NISQ) era \cite{preskill2018nisq}, in which actual quantum processing devices are made of a limited number of noisy qubits available in a number of different technological platforms. 
Restricting to the realm of digital quantum processors satisfying the DiVincenzo criteria \cite{DiVincenzo2000criteria}, typically 30 all-to-all connected qubits are available in trapped ions quantum processors \cite{Moses2023trapped_ions} and more than 100 in superconducting quantum circuits with next-neighbor coupling \cite{kim2023evidence}.
The physics of manybody quantum systems may thus be efficiently simulated through the direct application of such devices \cite{Tacchino2020AQT,daley2022review_qsim}. 
However, a proven advantage in quantum simulations is yet to be presented, at time of writing. In fact, while we may expect NISQ hardware to overcome known issues in simulating quantum systems on classical devices, such as the sign problem for fermionic systems \cite{abrams1997simulation,troyer2005computational}, a substantial limitation resides in the severely restricted hardware resources, nowadays. 
Nevertheless, 
first proof-of-concept quantum simulations of chemical properties have been reported \cite{kandala2017quantumchemistry,kandala2019mitigation,motta2022review}, as well as the time-dynamics of manybody Hamiltonians \cite{chiesa2019fourdimensional,arute2020observation,klymko2022evolution} and quantum field theories \cite{martinez2016real,klco2018quantum,de2022quantum,chakraborty2022classically,nguyen2022schwinger}.

In quantum simulations on NISQ devices, one of the most pressing targets is the efficient generation of a desired quantum state of a given model Hamiltonian on a quantum register, which is essential, e.g., to directly compute expectation values of the observables under study. This is the focus of the present work. 
We define $H_1$ as the target Hamiltonian, properly mapped on a quantum computer, and denote with $\{ |\Omega_n(1)\rangle  \}$ the set of its eigenstates. 
Using a sequence of single- and two-qubit gates \cite{nielsen00}, a unitary operator $U_n$ acting on a register of $L$ qubits such that $|\Omega_n(1)\rangle= U_n|0\rangle^{\otimes L}$ may always be found, and the number of gates required to exactly implement $U_n$ scales exponentially with $L$, i.e., with the size of the system. Currently, the noisy character of state-of-the art qubits and quantum gates available in NISQ devices limits the number of operations that can be reliably executed on actual hardware. Hence, significant effort has been devoted to propose efficient algorithms allowing to obtain a well approximated $U_n$ within limited quantum computing resources. For state preparation, variational quantum eigensolvers have been proposed as a valid solution, consisting of hybrid quantum/classical algorithms in which quantum circuits with parametrized operations are classically optimized to achieve the desired state \cite{peruzzo2014variational,McClean2016vqa_theory,moll2018optimization,anand2022quantum}.
Despite their usefulness for simple models, a general strategy to define such circuits for arbitrary systems is still not available; moreover, they may involve demanding classical optimizations in high-dimensional spaces \cite{McLean2018_barren}. An alternative approach has been recently proposed \cite{motta2020nphys}, based on imaginary time evolution, which might prove more effective than variational eigensolvers on NISQ devices, despite its intrinsic limitations in the preparation of quantum states with long correlation lengths.

In this work, we provide an original strategy to apply well-established methods developed in the context of Adiabatic Preparation (AP) \cite{born1928beweis} as a viable approach allowing to \textit{practically} achieve exponential advantage in quantum state preparation on quantum hardware. 
The latter allow to generate \emph{any} eigenstate of a given Hamiltonian, in the following referenced as the target Hamiltonian $H_1$, with high fidelity.
Specifically, the key idea is to start from a Hamiltonian $H_0$, different from $H_1$, whose eigenstates are analytically known and trivial to prepare on a quantum register. Then, its $n$-th eigenstate $|\Omega_n(0)\rangle$ is evolved in time from  $t=0$ to $t = \tau$, according to the auxiliary Hamiltonian
\begin{align}\label{eq:auxiliary hamiltonian}
    &\small{H\left( s\right) = \left(1 - f\left(s\right)\right)H_0 + f\left(s\right) H_1} \, , \quad s=t/\tau \, ,
\end{align} 
which is parametrized in terms of a dimensionless variable $s \in [0,1]$.
Essentially, $H\left( s\right)$ interpolates between $H_0$ and $H_1$ through the time-dependent holomorphic  function $f(s)$ such that $f(0) = 0$ and $f(1)=1$.
In what follows we denote the time evolution operator associated with $H(s)$ in the time interval $[0,\tau]$ as $U_\tau$. 
Its application to the n-th eigenstate of $H_0$ produces the final state $U_\tau \ket{\Omega_n(0)}$, which is an approximation of the actual eigenstate $\ket{\Omega_n (1)}$ of $H_1$, with a discrepancy quantified by the \textit{infidelity} measure, defined as 
\begin{equation}\label{eq: infidelity}
    \mathcal{I}(\tau) \equiv 1-\abs{\bra{\Omega_n(1)}U_\tau \ket{\Omega_n(0)}}^2 \, .
\end{equation}
In the limit $\tau \to \infty$, and provided that no level crossing occurs during the adiabatic evolution, the AP procedure becomes exact, i.e., $\mathcal{I}(\tau)\to 0$ and the final result is independent from the initial choice on $H_0$.
In practice, this is far from being the case, and the overall output infidelity 
strongly depends not only on the amount of finite available computational time $\tau$, but also on the specific choice of $H_0$ and the peculiar features of the the interpolating function, $f(s)$. In this respect, significant effort from the community went in the understanding of how different constraints on such quantities give rise to different upper bounds on the output infidelity \cite{albash2018adiabatic}.
Starting from different assumptions, several alternative bounds have been derived, most of which showcase a power-like dependence of the infidelity with the inverse of the characteristic adiabatic timescale $\tau$~\cite{ambainis2004elementary, elgart2012note, jansen2007bounds}.
Among these seminal theoretical works, in Ref.~\cite{nenciu1993linear} it is proven that the overall infidelity is bounded by an expression with an exponential rather than power-law decay as a function of $\tau$, for a broad class of interpolating functions. The latter, despite representing a theoretical milestone in the field of adiabatic processes, does not explicitly reveal the interplay between the infidelity's bound and the two arbitrary quantities $H_0$ and $f(s)$, thereby preventing its usage in a practical strategy for improving the performance of AP protocols.\\

The main goal of our work is to fill this gap. By revising the results provided in \cite{nenciu1993linear}, here we are able to determine the explicit form for an upper bound to the infidelity, which naturally accounts for the dependence on the spectral properties of the initial Hamiltonian $H_0$ and the interpolating function $f(s)$. 
Our result provides a practical protocol for selecting the $H_0$ and $f(s)$ that minimize the bound expression, thus leading to a potential significant improvement of the convergence performances of a standard AP algorithm run on actual NISQ hardware. 
In the following, we support the effectiveness of this protocol in choosing the initial $H_0$, by actually applying the AP approach to the determination of the ground-state configuration of two prototypical spin models, namely the Heisenberg and Ising models in either one or two dimensions. We thus perform numerical experiments at fixed quantum circuit depth, our numerical results demonstrate that our proposed scheme leads to a pronounced error suppression when compared to a standard AP protocol. 
These results are considerably relevant for prospective quantum simulations of many-body models on NISQ hardware with unprecedented system size, possibly towards quantum advantage.\\

\begin{figure}
    \centering
    \includegraphics{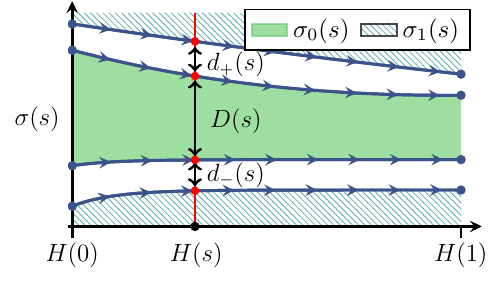}
    \caption{Schematic illustration of the adiabatic evolution of the spectral features related to a time-dependent Hamiltonian $H(s)$, in which $\sigma(s)$ is the full spectrum, partitioned into the subsets of eigenvalues $\sigma_0(s)$ and $\sigma_1(s)$. Here, $D(s)$ denotes the bandwidth of $\sigma_0(s)$, while $d(s) = \mathrm{min}\left[d_+(s),d_-(s)\right]$ defines the minimal gap between the two subsets of eigenvalues. }
    \label{fig: spectrum running}
\end{figure}

\textit{Exponential scaling and relevant parameters}.--
In the previous section we outlined AP ideas by focusing on the base case of determining a single eigenstate. We hereby pay attention to a more general scenario, and discuss our protocol in the case where the target of AP is a subspace, that is $\bar{V}$, of the entire Hilbert space $\mathcal{H}$. Specifically, let $\sigma_0(s)$ be a subset of the full spectrum $\sigma(s)$ of the interpolating Hamiltonian $H(s)$ defined in Eq.\ref{eq:auxiliary hamiltonian}. Furthermore, let us denote with $V(s)$ the subspace spanned by the eigenstates of $H(s)$ with eigenvalues in $\sigma_0(s)$. In particular, we assume that in our formalism $\bar{V}\equiv V(1)$. With this in mind, the AP task of targeting $\bar{V}$ is equivalent to devise a strategy for evolving $V(0)$ into $V(1)$ efficiently. In this respect, a proper figure of merit is given by the adiabatic error 
\begin{equation}
    \epsilon_{\mathrm{AP}}(\tau) \equiv \norm{\left( \mathds{1}-P(1)\right)U_\tau P(0)}\, ,
    \label{e:eAT}
\end{equation} 
where $\norm{A}=\mathrm{sup}_{\ket{\Psi} \in \mathcal{H}} \frac{\norm{A\ket {\Psi} }}{\norm{\ket {\Psi}}}$ denotes the sup(remum)-norm of the operator $A$, and with $\mathds{1}$ and $P(s)$ ($s\in [0,1]$) being the identity and the projector on $V(s)$, respectively.\\
It has been shown in \cite{nenciu1993linear} that under quite general conditions there exist a finite and positive $C$ and $g(s) \in \mathbb{R}$ such that, for $\tau$ large enough, the adiabatic error decays exponentially, 
that is
\begin{equation}\label{eq:errorAT}
\begin{split}
\epsilon_\mathrm{AP}(\tau) &\leq C \int_{0}^1ds\,\mathrm{exp}\left(- \left\lfloor \tau / g(s) \right\rfloor \right) \, ,
\end{split}
\end{equation}
in which $\left\lfloor \cdot \right\rfloor$ is the \textit{floor} operation. This is the case provided that $H(s)$ is holomorphic, that the gap $d(s)$ between $\sigma_0(s)$ and $\sigma_1(s) = \sigma(s) \symbol{92} \sigma_0(s)$ is positive, and that the bandwidth $D(s)$ of $\sigma_0(s)$ is finite for any $s \in [0,1]$ (see e.g. the scheme displayed in Fig.~\ref{fig: spectrum running}).
By defining the characteristic decay time as 
\begin{equation}
    \Tilde{g} \geq \mathrm{sup}_{s\in[0,1]}g(s) \, ,
\end{equation}
the exponential bound 
\begin{equation}\label{eq:exponential bound}
    \epsilon_\mathrm{AP} \leq C \,\mathrm{exp}\left(- \left\lfloor \tau / \Tilde{g} \right\rfloor \right) 
\end{equation}
is straightforwardly obtained, which clearly shows that a reduction of $\Tilde{g}$ determines 
an exponential reduction of $\epsilon_\mathrm{AP}$. \
In this work we extend the results of \cite{nenciu1993linear} where only the existence of $g(s)$ is proven, by finding an analytic explicit expression for $g(s)$ that relates it to the metrics $\{\norm{\Delta}\equiv \norm{H_0-H_1}, D(s),d(s)\}$, thus providing a concrete way to control $\Tilde{g}$ and therefore to define an efficient optimization strategy. In addition, it is relevant to notice that these metrics can be efficiently estimated by classical means for systems that are not too large, e.g., via the Lanczos Algorithm \cite{lanczos1950iteration, 2020SciPy-NMeth},
with a computational cost that is $\mathcal{O}\left( \mathrm{dim}\mathcal{H}\right)$. 
In the Supplemental Material (SM)
we report a possible approach to tackle the reduction of $\Tilde{g}$ also for large systems.\\
As we explicitly derive in the SM, a possible choice for $g(s)$ reads
\begin{equation}\label{eq:g_expression}
    g(s) = 4 \frac{W\left(d(s), D(s) \right)}{\varrho\left( s, \frac{d(s)}{ \norm{\Delta}}\right)}
\end{equation}
where $W(x, y) = \left[ 2^8 \left(1+\frac{2y}{\pi x}\right)^2 \left(\frac{\pi^2}{3} + \frac{4}{x} \right)\right]^{7/3}$
and with $\varrho\left(x,y \right) = \mathrm{sup}_{\rho \in \mathds{C}}\left[|\rho|\,  \mbox{such that}\,
    \left|f(x+ \rho)-f(x)\right| < \frac{y}{4}\right]$.
A direct inspection of Eq.~\eqref{eq:g_expression} reveals that $g(s)$ is a non-increasing function of $d(s)$ and $1/D(s)$. As a consequence, one obtains that 
\begin{equation}
    \Tilde{g} = 4\frac{W(d_\mathrm{min}, D_\mathrm{max})}{\mathrm{inf}_{s \in [0,1]}\varrho(s, \frac{d_\mathrm{min}}{ \norm{\Delta}})}\,,
\end{equation}
in which we defined the minimum of $d(s)$ and the maximum of $D(s)$ as $d_\mathrm{min} = \mathrm{inf}_{s\in[0,1]}d(s)$ and $D_\mathrm{max} = \mathrm{sup}_{s\in[0,1]}D(s)$, respectively.
For the commonly employed linear holomorphic function $f(s)=s$, the expression above reduces to
\begin{equation}\label{eq: g for linear thermalization}
    \Tilde{g} = 16\frac{\norm{\Delta}}{d_\mathrm{min}}W(d_\mathrm{min}, D_\mathrm{max}) \,.
\end{equation}

Notice that a dependence of the bound on $\norm{\Delta}$ has already been shown in the literature~\cite{jansen2007bounds}.
However, in all previous results the bounds resulted in an inverse power dependence on $\tau$. Not only our result confirms that a smaller $\norm{\Delta}$ is preferable,
but it also ensures that its reduction provides an exponential advantage.\\

\emph{Selecting $H_0$}.-- In the context of AP, given a target Hamiltonian $H_1$, the naive choice for $H_0$ is the diagonal part of $H_1$ (see, e.g., Refs.~\cite{mc2024towards, granet2023continuous}), such that its eigenstates are manifest.
In what follows, we present a universal procedure to select a more convenient auxiliary Hamiltonian, $H_0$, in the context of digital quantum simulations, where $U_\tau$ is constructed using the \textit{Trotter-Suzuki} approximation  \cite{trotter1959product, gauss1814methodus, quarteroni2006scientific}. 
Our proposal may not lead to the absolute best choice for specific problems, but it guarantees an exponential advantage, in the quantum state preparation of a target Hamoltonian $H_1$, over the naively picked $H_0$. 
The most generic starting Hamiltonian can be expressed in terms of its matrix elements with respect to the computational basis as 
\begin{equation}\label{eq: H_0 form}
   \bra{i}H_0\ket{j} \equiv \delta_{ij}\bra{i}H_1\ket{j}+\bra{i}M\ket{j} \, ,
\end{equation}
in which $M$ denotes a parameterized Hermitian operator (called the \textit{preconditioner})
to be optimized in order to minimize $\Tilde{g}$.
We impose three conditions on $M$:
(i) $H_0$ should be diagonal, such that its eigenstates are the elements of the computational basis and it is then trivial to initialize   them on a quantum computing register; (ii) the additional computational cost for the implementation of  $\mathcal{U}_M(\theta) = \mathrm{exp}\left( -i \theta M \right)$ at each Trotter step (where $\theta \in \mathbf{R}$) should not frustrate the cost reduction stemming from the minimization of $\Tilde{g}$; finally, (iii) the number of real parameters needed to describe $M$ should not increase exponentially with the number of qubits, $L$.
The first condition is satisfied by any diagonal operator. We hereby propose the extremely ``cheap'' form
\begin{equation}\label{eq:recipeM}
       \bra{i}M\ket{i} =  \sum_{j=0}^{L-1}\alpha_j\,\mathrm{bin}_j \left(i \right) \, ,
\end{equation}
in which $\ket{i}$ is the i-th element of the computational basis, and $\mathrm{bin}_j \left(i \right)$ is the j-th digit of $i$ once expressed in binary notation. The key idea behind the choice of Eq.~\eqref{eq:recipeM} is detailed in the SM. Most importantly, this choice allows to fulfill both conditions (ii) and (iii). In fact, the number of free parameters is $L$ in this case, thus increasing linearly with the system size. Moreover, the corresponding time evolution operator is
\begin{equation}\label{eq:UMcircuit}
   \mathcal{U}_M(\theta) = \bigotimes_{j=0}^{L-1}\mathrm{P}_j (\alpha_j \theta ) \, ,
\end{equation}
in which $\mathrm{P}_i$ is the single-qubit phase gate, applied to the i-th qubit. Notice that the quantum circuit depth of $\mathcal{U}_M(\theta)$ corresponds to a single qubit gate, independently from the size of the system.
Essentially, we propose to interpolate between $H_0$ (Eqs.~\eqref{eq: H_0 form}-\eqref{eq:recipeM}) and the targeted $H_1$ by AP, after classically minimizing $\Tilde{g}$ through a scanning of the free parameters $\{ \alpha_j\}$ in $H_0$. This simple protocol will guarantee an exponentially faster quantum state preparation.
Furthermore, since the approach we propose 
is independent from the realization of the Trotter evolution on the quantum register,
it can be fully combined with other optimization techniques such as those in Refs.~\cite{kovalsky2023self,perez2021quantum,mc2024towards}.

{\textit{Numerical results}}.--
Here we apply the AP to the preparation of the ground state for the Heisenberg model in either one- or two-dimensional (1D, 2D) square lattices with periodic boundary conditions. The purpose is to numerically confirm the previous theoretical result in case models of practical relevance for quantum simulations. 
The corresponding quantum circuits are implemented and executed on the Qiskit \textit{qasm} simulator \cite{Qiskit}. 

To focus only on $\epsilon_\mathrm{AP}$, we use sufficiently short Trotter steps, to suppress the systematic error coming from the Trotter-Suzuki approximation.
In the following, the state preparation is performed by applying three different approaches, all with the same adiabatic function $f(s) = s$. The first is the standard AP, 
where $M$ is set to zero in Eq.~\eqref{eq: H_0 form}.
Since the calculation of $\norm{\Delta}$ 
comes at the computational cost of a single application of the Lanczos algorithm (or an equivalent one), and it is independent from the chosen target subspace $V(s)$, we consider a second approach in which the parameters $\alpha_j$ are chosen to minimize $\norm{\Delta}$ in Eq.~\eqref{eq: g for linear thermalization}. Notice that this does not guarantee that $\Tilde{g}$ is minimized. In fact, we finally consider a third approach consisting in the full classical evaluation of $\Tilde{g}$ in Eq.~\eqref{eq: g for linear thermalization} involving the estimation $d_\mathrm{min}$ and $D_\mathrm{max}$. Despite yielding a significantly better scaling for our case studies, for more complicated models it may be less practical, since this approach requires the knowledge of the spectrum at several intermediate values of $s$, each requiring a Lanczos algorithm application.\\

\begin{figure}[t]
    \begin{subfigure}[]{0.48\textwidth}
        \includegraphics{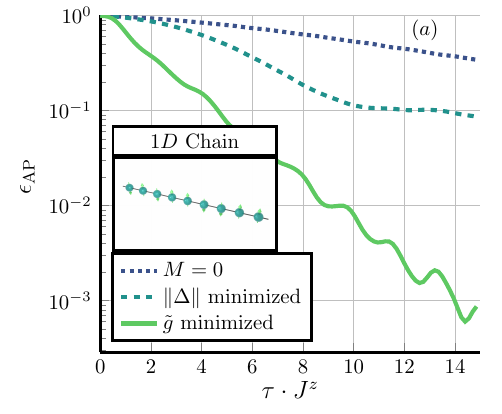}
    \vspace{-1.07\baselineskip}
        \includegraphics{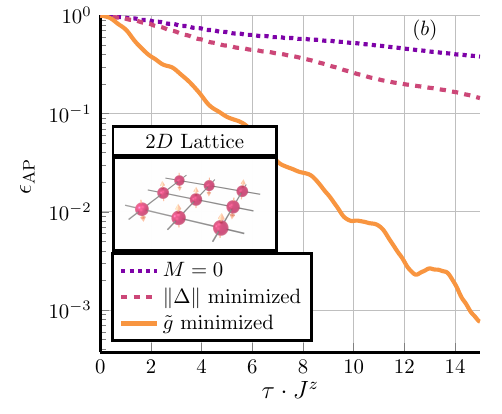}
    \end{subfigure}
    \caption{Ground state preparation for the 9 sites 1D  and 2D Heisenberg models
    with periodic boundary conditions, schematically illustrated in panels (a) and  (b), respectively, where the error dependence is plotted as a function of preparation timescale, $\tau$, for $J^x = 5J^z$.}
    \label{fig: H}
\end{figure}

\begin{figure}[t]
\includegraphics{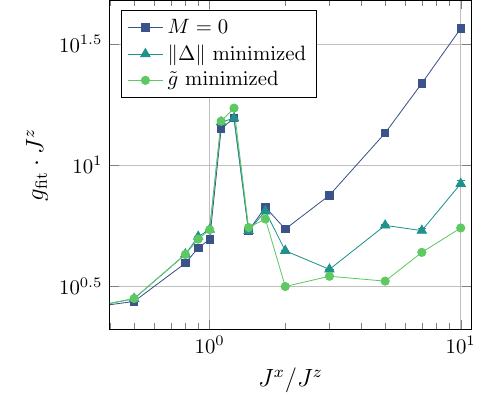}
\caption{Characteristic times $g_\mathrm{fit}$
obtained from fitting the numerical curves $\epsilon_\mathrm{AP}(\tau)$ (as the ones in Fig.~2) with a single exponential (see text), for different values of the ratio $J^x/J^z$ of the 1D Heisenberg model. Standard deviations from the fit are plotted as error bars (whose size is sometimes smaller than the markers).}
\label{fig: gfit}
\end{figure}

The XZ Heisenberg model is a prototypical manybody model describing systems of spin-1/2 particles on a lattice, defined by the interacting Hamiltonian
\begin{equation}
    H_1 = -\frac{1}{2}\sum_{\langle i,j \rangle}\left( J^z Z_i Z_j + J^x X_i X_j\right),
\end{equation}
where the sum runs over next-neighboring sites,
while $X_i$ and $Z_i$ are the Pauli matrices acting on the i-th qubit, conventionally defined as  
\begin{equation}
    X = \begin{bmatrix}
        0&1\\
        1&0\\
    \end{bmatrix}\,,
    \quad
    Y = \begin{bmatrix}
        0&-i\\
        i&0\\
    \end{bmatrix}\,,
    \quad
    Z = \begin{bmatrix}
        1&0\\
        0&-1\\
    \end{bmatrix} \, .
\end{equation}
In this work we consider real and positive coupling constants, $J^z$ and $J^x$. The AP process is then described by the operator 
\begin{equation}
    H(s) = -\frac{1}{2}\sum_{\langle i,j \rangle}\left( J^z Z_i Z_j + sJ^x X_i X_j\right)+ (1-s)M \, .
\end{equation}
At $s=0$ the Hamiltonian is diagonal and, 
in the unpreconditioned case ($M = 0$), 
has a degenerate ground state. The corresponding eigenstates, $\ket{0}^{\otimes L}$ and  $\ket{1}^{\otimes L}$, are orthonormal and invariant under translations.
It is possible to preserve the translational invariance of the AP process by imposing the condition $\alpha_i = \alpha$ ($\forall$ $i$) in Eq.~\eqref{eq:UMcircuit}.
This choice, albeit not unique, allows to reduce the number of parameters to a single one. 
The effect of the preconditioner ($M$) on the physical system is equivalent to the introduction 
of an external uniform magnetic field along the 
$z$ axis, with field intensity $\alpha$. 
The results for the ground state preparation of this model are reported in Fig.~\ref{fig: H}, 
for either the 1D or the 2D model. 
The outcome is quite striking: independently on the dimensionality of the model or the parameters regime, the AP provides a way of determining the ground state of the XZ Heisenberg model with an error that scales almost exponentially with the preparation timescale. In addition, by comparing the three different approaches above, we notice a significant reduction in the required preparation timescale to achieve a given error, with the third (most optimal) approach allowing to obtain the shortest $\tau$. The performance differences between the different approaches become more and more evident on increasing $J^x/J^z$, as shown in Fig.~\ref{fig: gfit}. Here, the characteristic time obtained by fitting ``$\ln{\{\epsilon_\mathrm{AP}(\tau)\}}$'' with the function ``$C_\mathrm{fit}- \tau/g_\mathrm{fit}$'' is reported as a function of $J^x/J^z$.
In particular, the $g_\mathrm{fit}$ values
obtained with the three approaches 
are nearly indistinguishable when $J^x \ll J^z$, 
while they exhibit a clear advantage after applying
our preconditioning strategy for $J^x \gg J^z$,
i.e. when the target Hamiltonian 
is significantly different from the starting one.
The peak in the plot occurring around $J^x/J^z \simeq 1$ corresponds to the point where, for an infinite number of spins, the system would exhibit a phase transition. In that region the preparation of the ground state is still a rather expensive task and our proposal is less effective. 
Similar results have been obtained for the Ising model in a transverse magnetic field, in both one and two dimensions, as reported in the SM for completeness. 
In practice, the AP of excited states would be more involved, due to the increased dimension of the sub-spaces involved in the time evolution. However, different techniques could be exploited to this purpose, which we aim to explore and communicate in a future work.
As a final comment we notice that, as far as these very specific spin models are concerned,
the actual improvement due to the preconditioning term only occurs for an odd number of sites with periodic boundary conditions, for which a detailed explanation is given in the SM.

{\textit{Conclusions.--}
{We have applied the theory of Adiabatic Preparation (AP) to the framework of quantum simulations on NISQ devices, specifically focusing on the efficient preparation of a given quantum state. 
Starting from the seminal result of Nenciu
in~Ref.~\cite{nenciu1993linear}, 
we have analytically derived an exponential bound for the AP error, 
thus obtaining for the first time
an explicit expression for its characteristic time,
in terms of quantities that can be classically estimated. 
In future works, it would be beneficial to examine the tightness of such a bound with analytical evidence or numerical tests.
Using this improved analytical control, we propose a universal preconditioning term, denoted as $M$ and tailored for NISQ computing \cite{mckay2017efficient}, which can be effectively optimized using the aforementioned formulation. 
Upon incorporation into the Hamiltonian model to be simulated on quantum hardware, $M$ allows to exponentially suppress the AP error.
Finally, we have tested our protocol by numerical simulations performed for the representative Heisenberg and Ising models, in either one- or two-dimensional dimensions, respectively.
Our results demonstrate an exponential improvement, particularly notable under more strenuous computational scenarios. Despite performing numerical tests on simple spin models, we expect similar behaviours to occur for more complicated systems. We envision that our work will significantly enhance the efficiency of quantum state preparation, thereby boosting the prospective achievement of quantum utility in state-of-the-art digital quantum devices. }

{\textit{Acknowledgements.--}
This research was supported by the Italian Ministry of Research (MUR): D.G. acknowledges the PNRR project CN00000013 - National Research Center on ``HPC, Big Data and Quantum Computing'' (HPC), D.N. acknowledges the PNRR project PE0000023 - National Quantum Science Technology Institute (NQSTI). The research of M.B. for this work was partially funded through the MUR program for young researchers “Rita Levi Montalcini”. M.B. is (partially) supported by ICSC - Centro Nazionale di Ricerca in High Performance Computing, Big Data and Quantum Computing, funded by European Union – NextGenerationEU.
The authors warmly acknowledge G. Guarnieri, F. Scala, and F. Tacchino for several scientific discussions and suggestions.

\bibliography{main.bib}

\end{document}